\documentclass[fleqn]{aa}  

\usepackage{graphicx}
\usepackage{xcolor}
\usepackage{hyperref}
\usepackage{txfonts}
\usepackage{calrsfs}
\usepackage[T1]{fontenc}
\usepackage{tabularx}
\usepackage{siunitx}
\usepackage[mathcal]{eucal}
\usepackage{longtable}
\usepackage{multirow}
\usepackage{adjustbox}
\DeclareSIUnit \parsec {pc} 

\begin{document} 

\title{Parameter-free Hubble constant from the quadruply lensed quasar SDSS J$1004+4112$}
\author{Joseba Mart\'inez-Arrizabalaga\inst{1}\fnmsep\thanks{joseba.martinez@alumnos.unican.es} 
  \and
  Jose M. Diego\inst{2}
  \and
  Luis J. Goicoechea\inst{1}}
 \institute{Universidad de Cantabria, Avda. de Los Castros 48. 39005 Santander, Spain
 \and
 Instituto de F\'isica de Cantabria (CSIC-UC). Avda. Los Castros s/n. 39005 Santander, Spain
    }     
\providecommand{\keywords}[1]
{
  \small        
  \textbf{\textit{Keywords---}} #1
}
\date{Received 25 September 2023 / Accepted 4 December 2023
}
\abstract{We present a free-form lens model for the multiply lensed quasar in the galaxy cluster SDSS J$1004+4112$. Our lens model draws minimal assumptions on the distribution of mass in the lens plane. We have paid particular attention to the model uncertainties on the predicted time delay originating from the particular configuration of model variables. Taking into account this uncertainty, we obtained a value of the Hubble constant of $H_0= 74^{+9}_{-13}\SI{}{\kilo\meter\per\second\per\mega\parsec}$, which is consistent with recent independent  estimates. The predicted time delay between the central image E and image C (the first to arrive) is $\Delta T_{E-C}=3200\pm 200$ days for the estimated Hubble constant. Future measurements of $\Delta T_{E-C}$ will allow for a tighter constraint to be imposed on $H_0$ in this cluster-QSO system. 
}

\maketitle

\section{Introduction}\label{sec_intro}

The SDSS J1004+4112 galaxy cluster is a gravitational lens at a redshift of $z = 0.68$, which produces five images of a background quasar at redshift $z = 1.734$. This gravitationally lensed quasar was found in the Sloan Digital Sky Survey by \citet {inada2003gravitationally}, when the large separation between its multiple images was noticed. Following the discovery of SDSS J$1029+2623$ \citep{Inada_2006}, currently SDSS J1004+4112 is the second lensed quasar with the greatest reported separation. 

There is a broad range of previously published works based on this lens system and many of them have been focused on observational constraints, for instance, cluster substructure \citep{oguri2004observations}, central quasar image \citep[faintest image E;][]{naohisa2005discovery}, background lensed sources
\citep[][]{sharon2005discovery,liesenborgs2009non, Oguri_2010}, and time delays between the four brightest images (ABCD) of the quasar \citep{fohlmeister2007time,fohlmeister2008rewards,munoz2022longest}.

Additionally, most lens mass models of SDSS J1004+4112 were performed using parametric techniques: \citet{inada2003gravitationally,oguri2004observations,fohlmeister2007time,fohlmeister2008rewards,Oguri_2010,For_s_Toribio_2022,napier2023hubble,liu2023hubble}. Some non-parametric models are also available \citep[e.g.,][]{williams2004models,Saha_2006,liesenborgs2009non,mohammed2015lensing}. However, the model derived by \citet{perera2023freeform} is the only free-form model that includes the three independent delays between the four brightest images that have been recently and accurately measured \citep{munoz2022longest}.

Most recently, following the method proposed by \cite{refsdal1964possibility}, \cite{napier2023hubble} used the \texttt{Lenstool} software to measure the Hubble constant from six measured time delays (including three independent delays of SDSS J1004+4112) and parametric mass models. They obtained $H_0 = \SI[separate-uncertainty = true]{71.5(61)}{\kilo\meter\per\second\per\mega\parsec}$. In addition, \cite{liu2023hubble} focused on SDSS J1004+4112, using the \texttt{GLAFIC} software to estimate $H_0$ from its measured delays and 16 parametric models with equal weight. The resulting Hubble constant is $H_0=67.5^{+14.5}_{-8.9}\SI{}{\kilo\meter\per\second\per\mega\parsec}$.

In this work, we explore Refsdal's method in the lensed quasar with the most precise time delays among those with a large number of observation constraints for the lensing mass and image separation exceeding $10/$arcsec. A large number of observation constraints are required to successfully reconstruct the lensing mass from a non-parametric method, so galaxy-scale lensed quasars are usually studied from parametric models. At present, there are no non-parametric lens mass reconstructions that use the three well-measured time delays of SDSS J1004+4112 to estimate the Hubble constant. Thus, the current work complements the two recent studies of the system \citep[see:][]{liu2023hubble,napier2023hubble} to achieve a more comprehensive perspective. 

For the lens mass reconstructions, the redshifts and image positions of seven lensed galaxies were considered, taken from \citet{Oguri_2010}. The lensed sources are found at four different redshifts, which means that we do not need to be concerned about the mass-sheet degeneracy \citep{Diego2007}. In the case of sources at different redshifts, different models tend to find similar mass profiles that are not biased, demonstrating that the degeneracy does not exist for multiple lensed sources \citep{2017_degeneracybreak}. With regard to the measured time delays, as in the work of \citet{For_s_Toribio_2022}, their formal errors  \citep[see ][]{munoz2022longest} are enlarged by a factor of 5. We also remark that quasar image flux ratios at near-infrared and radio wavelengths (which are less sensitive to microlensing effects) are used to weight mass models. The method used for this cluster can be applied to future data with similar measurements. 

The paper is organized as follows: Section 2 gives a brief description of the gravitational lensing theory, proposing the mathematical formulation of the problem and briefly explains the WSLAP+ strong lensing inversion code \citep{Diego_2005,Diego2007} used to derive the free-form lens model. Section 3 presents 21 different reconstructed lens models, used to account for the uncertainty in the predicted time delays from the lens model. In Section 4, the likelihood functions of $H_0$ and the weight of each model are discussed. Finally, in Section 5, the obtained estimation of $H_0$ is shown, followed by a discussion of the results in Section 6.

In this work, we assume a cosmology with parameters $\Omega_M = 0.3$ and $\Omega_\Lambda=0.7$. The mass models derived in this work scale as $1/H_0$, which depends very weakly on the values of $\Omega_M$ and $\Omega_\Lambda$ within the currently accepted ranges of these parameters. The time delay also scales as $1/H_0$ which is by far the most relevant cosmological dependence.


\section{Lens modeling}\label{sec_lensing}
The reconstructed lens models of SDSS J$1004+4112$ described in this article are obtained with the free-form code WSLAP+ \citep{Diego_2005,Diego2007}. Details are given in the references. Here, we simply give a brief description of the algorithm.
The lensing problem in its most basic form is formulated linearly as
\begin{equation}
    \label{eq:LensEq. }
    \vec{\beta} = \vec{\theta}- \vec{\alpha}(\vec{\theta},M(\vec{\theta})),
\end{equation}
with $\vec{\theta}$ corresponding to the observed positions of the lensed images, $\vec{\beta}$ the unknown  positions of the background galaxies, $M(\vec{\theta})$ the unknown mass distribution of the lens and $\vec{\alpha}(\vec{\theta},M(\vec{\theta})$ the deflection angle that corresponds to the observed position $\vec{\theta}$. 

The deflection angle is the contribution to the deflection of all the mass distribution, that is obtained by integrating 
\begin{equation}
    \label{eq:deflection_exact}
    \alpha (\theta,M(\vec{\theta})) = \frac{4G}{c^2}\frac{D_{ls}}{D_sD_l}\int M(\theta ') \frac{\theta - \theta '}{|\theta -\theta '|^2}d\theta '
\end{equation}
where $D_{ls}$, $D_{l}$, and $D_s$ are the angular distances from the lens to the source galaxy, from the observer to the lens, and from the observer to the source correspondingly.  

The problem is discretized assuming the lens is divided into $N_c$ cells, in each of which the mass is nearly constant, assigning a mass $m_i$ to each cell. The deflection angle (\ref{eq:deflection_exact}) is then approximated by a sum 
\begin{equation}
    \alpha(\theta) =\frac{4G}{c^2}\frac{D_{ls}}{D_sD_l}\sum _{N_c} m_i \frac{\theta -\theta_i}{|\theta -\theta_i|^2} = \gamma M
\end{equation}
this is, the deflection is approximated by the contribution of $N_c$ discrete masses $m_i$ (contained in the array of masses $M$), in positions $\theta_i$ in the direction of $\theta-\theta_i$ and magnitude $m_i/|\theta-\theta_i|$. The matrix $\gamma$ hence contains the deflection field at position $\theta_j$ from a grid point at position $\theta_i$ with mass $m_i=1$. 


The galaxy cluster SDSS J$1004+4112$ shows a distribution of tangential strong lensing arcs over the image. Discretizing these lensed arc positions into $N_\theta$ points, the lens equation  can be formulated as a linear problem between unknown positions $\beta$ and the masses $m_i$

\begin{equation}
    \beta = \theta -\gamma M
    \label{Eq_BetaM}
\end{equation}
where $\beta$ and $\theta$ are $2N_\theta$ vectors containing the x and y unknown positions in the source plane and the image positions in the image plane, while $M$ is the mass vector containing the $N_c$ mass cells. $\gamma$ is a $2N_\theta \times N_c$ whose i-th row $\gamma_i$ gives the linear relation between the mass distribution contained in the vector $M$ and the deflection angle in the corresponding $\theta_i$ position contained in the i-th row of the $\theta$ vector. 
The mass distribution can also contain a contribution from a compact component associated with member galaxies. This component can be divided into different layers that from the point of view of Eq. ~\eqref{Eq_BetaM} behave as additional grid points. The number of layers must be at least $N_l=1$ if all galaxies are placed in the same layer and up to  $N_l=N_{gal}$ if we assign one layer per galaxy. In this work, we adopt one single layer for all galaxies, that is, $N_l=1$. 

The system has $2N_s + N_c + N_l $ unknowns (lens masses and source positions), that can be reduced assuming the $N_s$ sources can be approximated by point sources, this is, forcing the $\theta$ positions of the arcs correspond to the $N_s$ source positions. This can be formalized as
\begin{equation}
    \label{eq:Lensing5Matrix}
    \theta = \Gamma X,
\end{equation}

where $\Gamma$ is a $2N_\theta \times (N_c + N_l + 2N_s)$ known matrix given by the grid configuaration and positions of lensed galaxies and $M$ contains all the unknowns, namely,\ the mass elements and positions of the lensed galaxies in the source plane. 

As the problem may be ill-conditioned and too big to use matrix inversions, approximate numerical methods are used, 
for instance, the bi-conjugate gradient method \citep{press1997numerical} can produce a fast solution to the system of equations given by Eq. ~\eqref{eq:Lensing5Matrix}. 
However, models obtained with this approach often return unphysical solutions where elements in the mass vector, $M,$ can adopt negative values. This is solved by imposing the constraint that the mass must be always positive. This type of problem can be solved by somewhat slower, but still very fast quadratic programming algorithms \citep[see][]{Diego_2005}. 

\begin{figure}[h]
\centering
\includegraphics[width=0.8\linewidth]{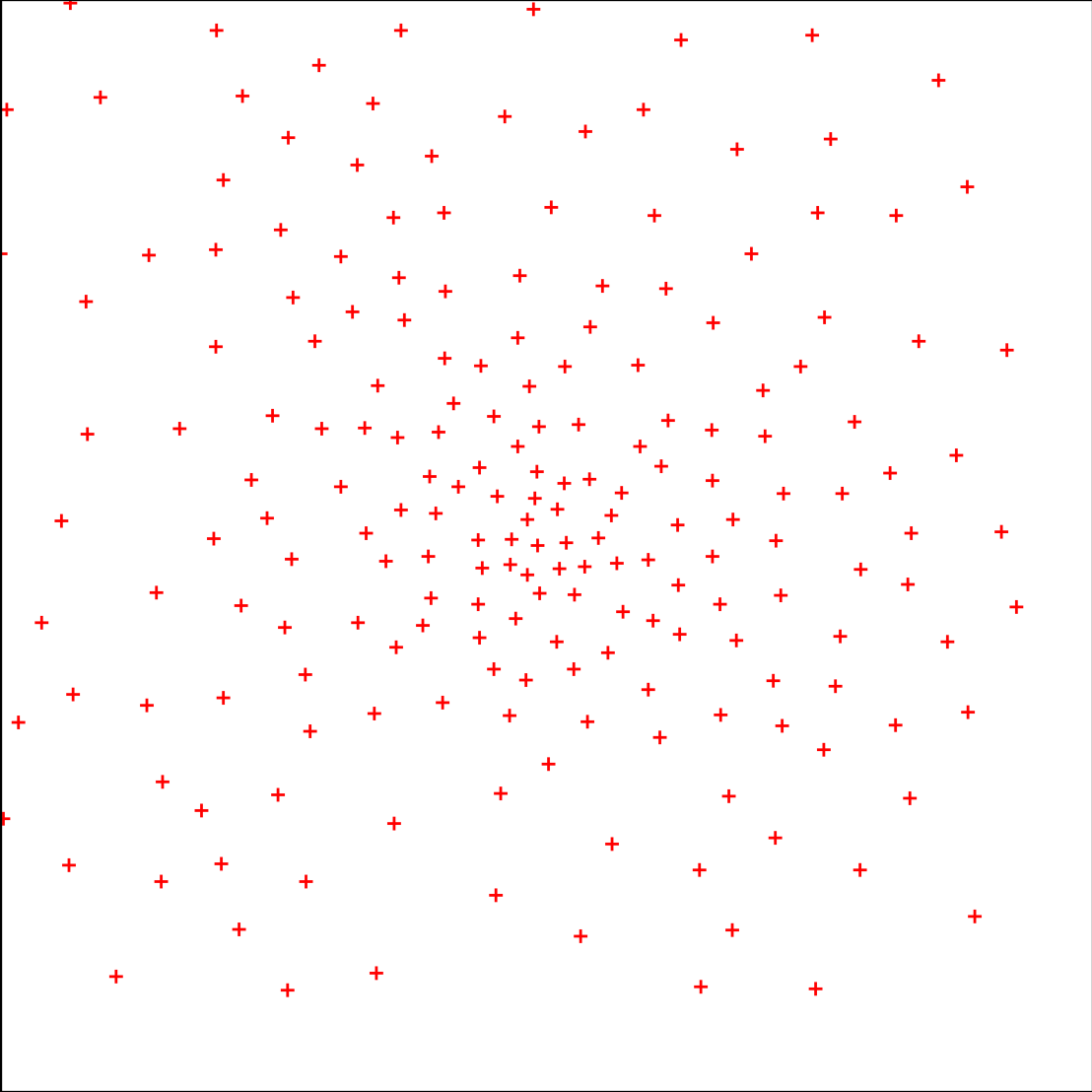}
\caption{Example of a dynamic grid with 199 points used for the smooth dark matter distribution.}
\label{fig:199_DynamicGrid}
\end{figure}

A first solution is obtained where the lens plane is initially discretized with a regular $16\times 16$ grid. This is similar to assuming no prior knowledge about the distribution of the mass since all grid points play the same role in the system of equations given by Eq. \eqref{eq:Lensing5Matrix}. After this first solution, a multi-resolution grid is performed, which is based on the previously obtained lens model. This new grid increases the number of grid points in the areas where the matter concentration of the original solution is greater by making the cells' size inversely proportional to the matter density. Fig. \ref{fig:199_DynamicGrid} shows a multi-resolution grid with 199 points, which concentrates more grid points in the center as a higher mass density is expected.

\section{Lens model results}\label{sec_results}
The uncertainty in the derived Hubble constant is dominated by the uncertainty in the lens model. In order to capture this uncertainty, we derived a range of models, all of them consistent with the observed set of linear constraints (positions of lensed galaxies and QSO). The observed time delays or relative magnifications, are not included in our set of constraints.  In the case of WSLAP+, a major source of uncertainty in the lens model originates in the particular choice of the grid used to describe the smooth component of the mass. The grid configuration is a free variable in WSLAP+. For the number of constraints used in this work, grid configurations with 150 to 300 points can produce satisfactory results in terms of reproducing the position of lensed galaxies with relatively smooth critical curves. A larger number of grid points results in unstable and unreliable solutions and smaller numbers result in a lack of resolution. The cases presented should capture the expected range of uncertainty. 

Another source of variability comes from the starting point of the minimization. It is possible to choose among an infinite number of initial conditions. After each minimization, each solution differs slightly from other solutions obtained with a different seed. We varied both the grid and initial seed to explore the range of valid models. \\

\subsection{Range of lens models}
We derived a total of 21 lens models: one with a uniform grid of $16 \times 16$ points, nine with a dynamical grid of 199 points, six with a different dynamical grid of 248 points, and an additional five models with a different dynamical grid of 299 points. However, only the 20 multiresolution grids were used to compute the estimation of the Hubble constant since the solution obtained with the regular 16x16 grid lacks resolution in the central region. 

\begin{figure}[h]
    \centering
    \includegraphics[width=\linewidth]{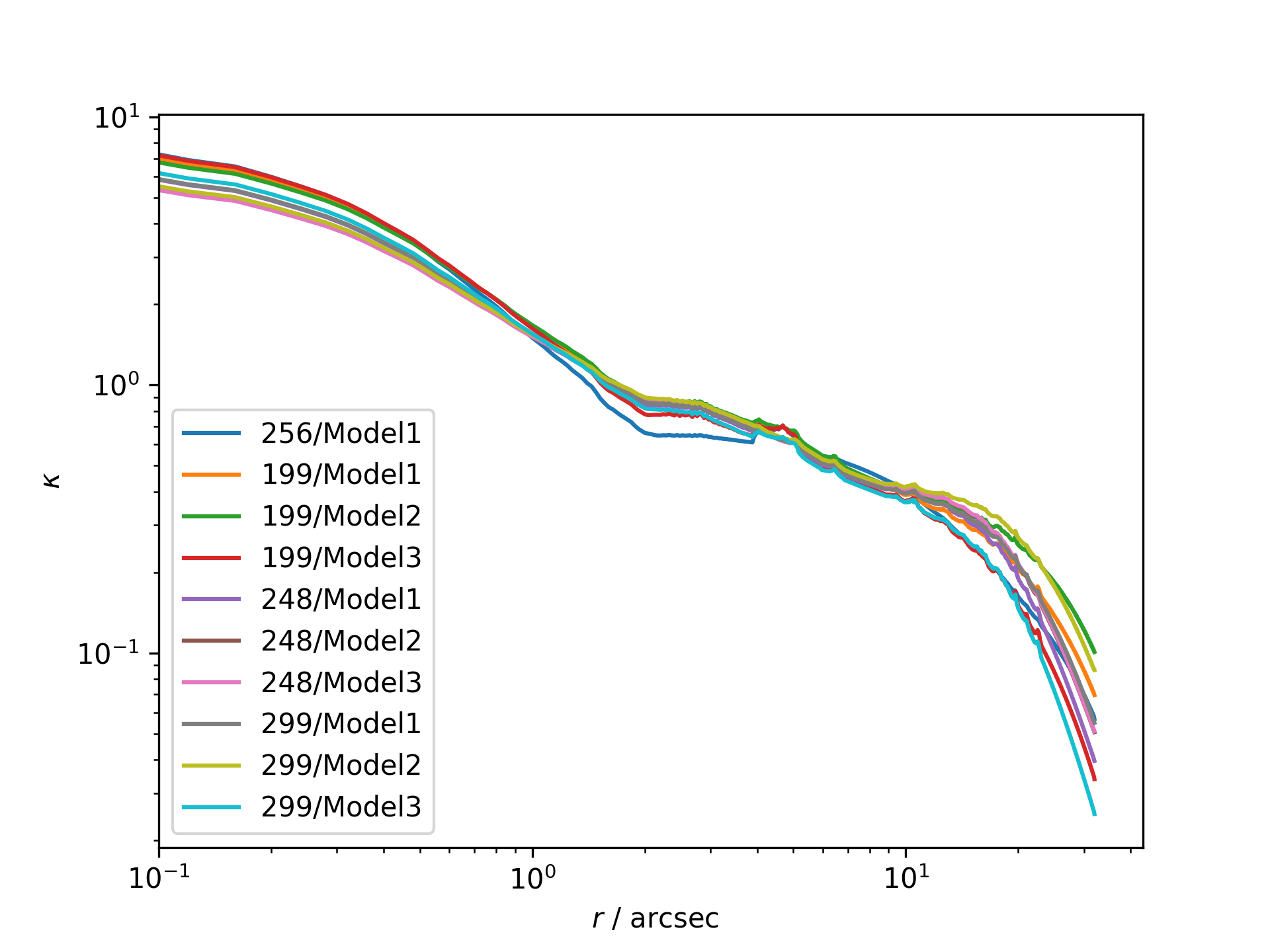}
    \caption{Convergence $\kappa$ for 10 models computed for a source at  $z=1$. }
    \label{fig:convergence}
\end{figure}

\begin{figure}[h]
     \centering
    \includegraphics[width=1\linewidth,trim={5cm 2cm 5cm 1.5cm},clip]{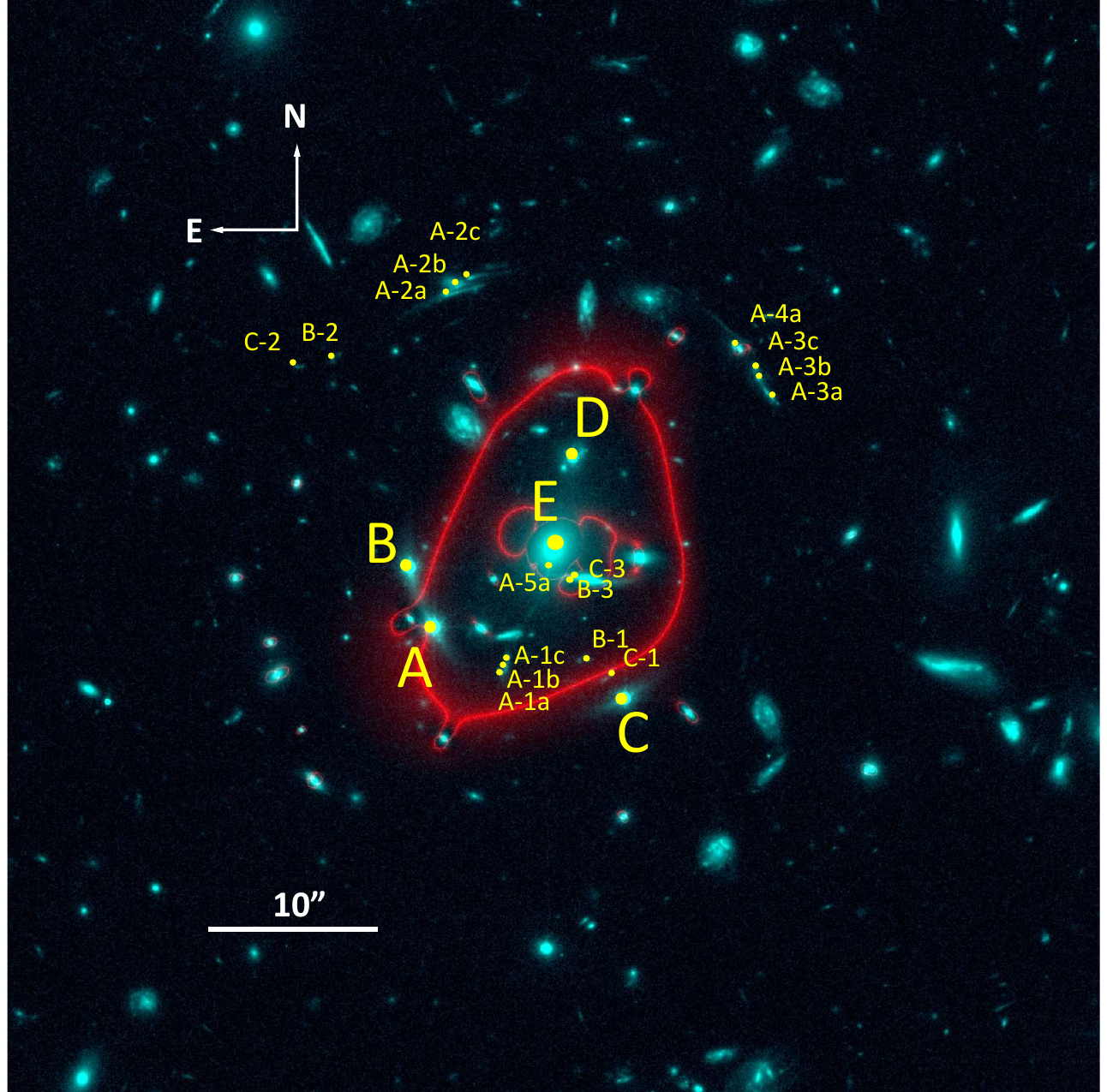}
\caption{Magnification map of the reconstructed lens of SDSS J$1004+4112$  using the dynamic grid with 199 points.}
\label{fig:caustics}
\end{figure}

Figure \ref{fig:convergence} shows the convergence of 10 of the 21 models used, including the initial model obtained with the uniform grid $16\times 16=256$ grid points.The first number in the legend indicates the number of grid points, while Model1, Model2, and Model3 are different realizations (different initial conditions) for the same grid configuration. Models that are not shown have similar profiles. These profiles are consistent with the ones derived by \cite{liu2023hubble} using a parametric model. Despite the relative similarity between profiles, the small changes between models are the main source of uncertainty in the predicted time delay, and hence in the derived value for the Hubble constant. Figure \ref{fig:caustics} shows the magnification map of the lens model,
obtained with the dynamic grid of 199 points in Fig. \ref{fig:199_DynamicGrid} plotted in red on top of the galaxy cluster shown in blue.

\subsection{Predicted time-delay for image E}
Each of the lens models described in the previous subsection gives a prediction for the image E time-delay for a canonical Hubble constant $H_0= 70\SI{}{\kilo\meter\per\second\per\mega\parsec}$. Thus, by re-scaling the predicted time delays to the value of the Hubble constant measured, $H_0= 74\SI{}{\kilo\meter\per\second\per\mega\parsec}$ (see Section 5), we can obtain a predicted time-delay with respect to image C of $3200\pm200$ days. This is approximately 1 year longer than the expected value reported in \cite{For_s_Toribio_2022}, who obtained a predicted time delay of about 8 years.

A robust determination of the time delay of image E (with respect to image C) would play an important role since this measurement would significantly improve the precision in the estimation of the Hubble constant. However, image E is very faint. This means that the typical magnification ratio between images D and E is $\mu_D/\mu_E \sim 30$ (with a great variation among different models) and, consequently, it is reasonable to assume that image E is about 30 times fainter than image D at optical wavelengths (neglecting possible extinction and microlensing effects).  Hence, if the $r$-band magnitude of D is $\sim$21 \citep[see Table 1 of][]{munoz2022longest}, the expected $r$-band magnitude of D would be $\sim$24.7 mag. Additionally, E is located in a bright and crowded sky region, so it seems extremely difficult to build its optical light curve from current facilities.


\section{Constraints on $H_0$}
It is common to express the Hubble constant $H_0$ as: $$H_0 = h \times \SI{100}{\kilo\meter\per\second\per\mega\parsec},$$ where $h=H_0/100$ is referred to as the reduced Hubble constant and it is a dimensionless physical constant. 

In order to measure $H_0$, \cite{refsdal1964possibility} proposed that the gravitational lensing effect could be used. In fact, when a light ray is deflected by a gravitational lens, its trajectory is modified and therefore its traveling time, due to the light's finite propagation speed. The expression that describes the time delay of the perturbed ray with respect to the one that follows a straight path is displayed in Eq. \eqref{eq:time_delay}: 

\begin{equation}
    \label{eq:time_delay}
    t(\vec\theta) =\frac{1+z_l}{c}\frac{D_lD_s}{D_{ls}}\left[\frac{1}{2}(\vec\theta-\vec\beta)^2-\Phi(\vec\theta)\right],
\end{equation}with
$z_l$ indicating the redshift of the lens.

The three angular distances involved in Eq.  \eqref{eq:time_delay} are inversely proportional to the Hubble constant, $H_0$, therefore, we say that the time delay scales as $1/h$. In addition, the magnitude expressed in Eq.  \eqref{eq:time_delay} is not measurable; instead, we chose to measure the time delay relative to one of the images.

Using the lens models derived in the previous section, we are able to confront the predictions of the time delay of the multiple images of the QSO with the observations, both relative to image C. Our lens models predict the time delay for a canonical value of $H_0=70$ km\,s$^{-1}$\,Mpc$^{-1}$, which is a reduced Hubble constant of $h=0.7$. Hence, these time delays are re-scaled to a new reduced Hubble constant (h') by multiplying the predicted time delays, $x=h'/0.7$

\subsection{Likelihood}
For each lens model, we can derive a probability for the presence of $h$ by minimizing the weighted difference between the predicted and observed time delays ($\Delta T_{M_i}$ and $\Delta T_{obs}$, respectively):
\begin{equation}
p(M_i,h) \propto \prod_{j=A,B,C,D} exp\left [  -\frac{(\Delta T_{M_i}^j-\Delta T_{obs}^j)^2}{2\sigma_j^2} \right ],
\label{Eq_P1}
\end{equation}
where the index J runs over the four positions, A, B, C, and D. In addition, the proportionality sign just expresses the lack of a normalization constant.
The values of $\sigma_j$  are computed at each of these four positions as the dispersions of the time delays predicted by the lens models. Image E has no observable time delay and hence is not included in the definition of the likelihood. 



Since we have different lens models derived under different assumptions, we need to compute the joint probability from all models. We adopted the conservative approach that the combined probability from all lens models is the sum of probabilities of the models \citep[see, e.g.,  ][]{Vega2018}. A more aggressive approach would be to consider that the joint probability is the product of the individual probabilities, but this implies all models are independent from each other. This is not true since all lens models are derived using the same algorithm, with some assumptions common to several models (e.g., the grid configuration, definition of the compact component, or lensing constraints). 

The joint probability is then defined as:
\begin{equation}
    P(M,h) \propto \sum_{i=1}^{N} w_i p(M_i,h),
    \label{Eq_P2}
\end{equation}
where the sum runs over the $N$ lens models, $p(M_i,h)$ is the probability given in Eq. ~\eqref{Eq_P1} and the term $w_i$ is the weight assigned to each model. Once again, the proportionality sign just indicates the lack of a normalization constant.  
This weight is determined by how well a particular model reproduces the observed flux ratios between the different images of the QSO. We define this weight as (dropping the subscript $i$); 
\begin{equation}
   w = \exp\left[-\frac{1}{2}\left(\frac{(\frac{F_B}{F_C}-\frac{\mu_B}{\mu_C})^2}{\sigma_{CB}^2} \right.\right. + 
                  \left.\left.\frac{(\frac{F_D}{F_C}-\frac{\mu_D}{\mu_C})^2}{\sigma_{CD}^2}\right)\right].
    \label{eq:weight}
\end{equation}
In the above definition, we have ignored image E since this is very close to the center of the BCG galaxy and extremely faint. In addition, initially, we did not include the flux ratio, $F_A/F_C$, in w Eq. \eqref{eq:weight} because the A image is very close to a critical line and, thus, the magnification ratio, $\mu_A/\mu_C$, in some models might be a strongly biased estimator of $F_A/F_C$. Hence, in an abundance of caution, image A was ignored for respect to the definition of the weight.
We computed the flux ratios as the average of three flux ratios from our own measurements in the IR band F814W, measurements published in \cite{Oguri_2010} and radio flux in \cite{hartley2021using}. The values of $\sigma_{CB}$, and $\sigma_{CD}$ are the combination of the observed dispersion of flux ratios between different measurements and the expected variation due to the intrinsic variability:
\begin{equation}
\sigma_{XY}^2 = \sigma_{XY_o}^2 + \sigma_{int}^2
.\end{equation}
For the intrinsic variability in the F814W HST filter, we adopted a value of $\sigma_{int}=0.2$ for image B and $\sigma_{int}=0.16$ for image D from \cite{Li2018} and \cite{Morganson_2014}. The measured time delays, flux ratios measured by the indicated authors, and estimated intrinsic variability are summarized in Table \ref{tab:summary}. The flux ratios from Oguri and this work correspond to the F814W HST filter and Hartley's in the 5GHz radio band.

\begin{table}[h]
    \centering
    \resizebox{\columnwidth}{!}{\begin{tabular}{cccccc}
    \hline
    \hline
        Image & A & B & C & D & E\\
        \hline
        Obs. delay (days) & $825.99\pm2.10$ & $781.92\pm 2.20$ & $\equiv 0 $ & $2456.99\pm 5.55$& -\\
        Flux (this paper)& $1.84\pm0.03$ & $1.50\pm 0.03$ & $\equiv 1$ &    $0.741\pm0.016$ & $0.015\pm 0.011$\\
        Flux (Oguri) & $2.2\pm0.6$ & $1.6\pm 0.65$ & $\equiv 1$ & $0.56\pm 0.22$ & $0.006\pm0.004$ \\
        Flux (Hartley) & $1.7\pm0.2$ & $1.10\pm0.17$ & $\equiv 1$ & $0.32\pm0.12$ & -\\      
        Int Variab. & 0.3 & 0.2 & 0 &0.16\\
        \hline
    \end{tabular}}
    \caption{Observed time delay, flux ratios, and adopted intrinsic flux ratio variation for the QSO images of SDSS J1004. }
    \label{tab:summary}
\end{table}



\section{\texorpdfstring{$H_0$}{TEXT} results}

\begin{figure}[h]
    \centering
    \includegraphics[width=\linewidth]{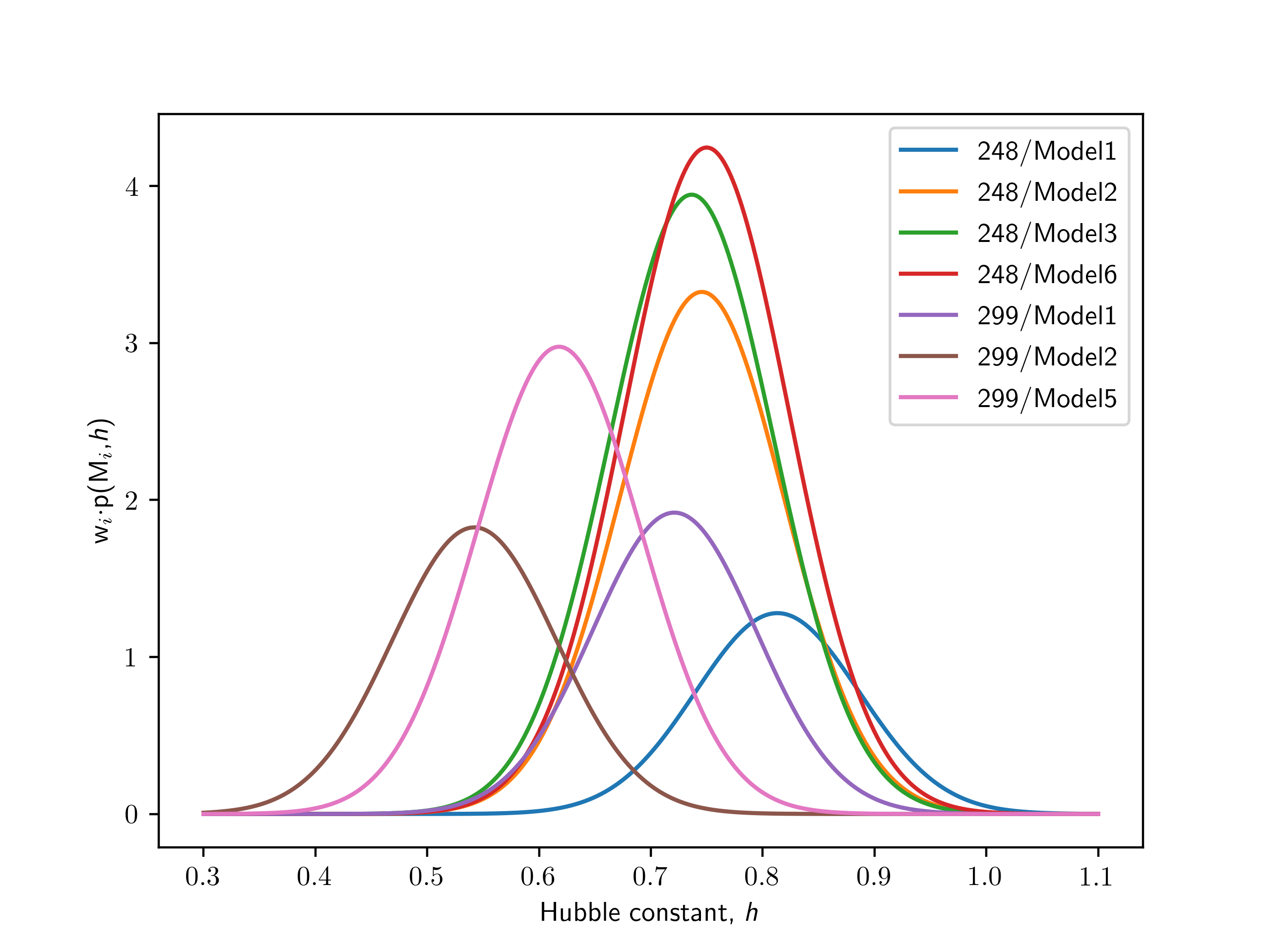}
    \caption{Contribution of the seven models with the highest weights to the final likelihood.}
    \label{fig:all_models_H0}
\end{figure}

Each of the models obtained in this work has a different contribution to the prediction of the value of the Hubble constant. {Figure \ref{fig:all_models_H0}} shows the contribution of each of the models to the final likelihood, this is, it shows the function defined by each weighted probability in the likelihood function in Eq. \eqref{Eq_P2}.

\begin{figure}[h]
    \centering
    \includegraphics[width=\linewidth]{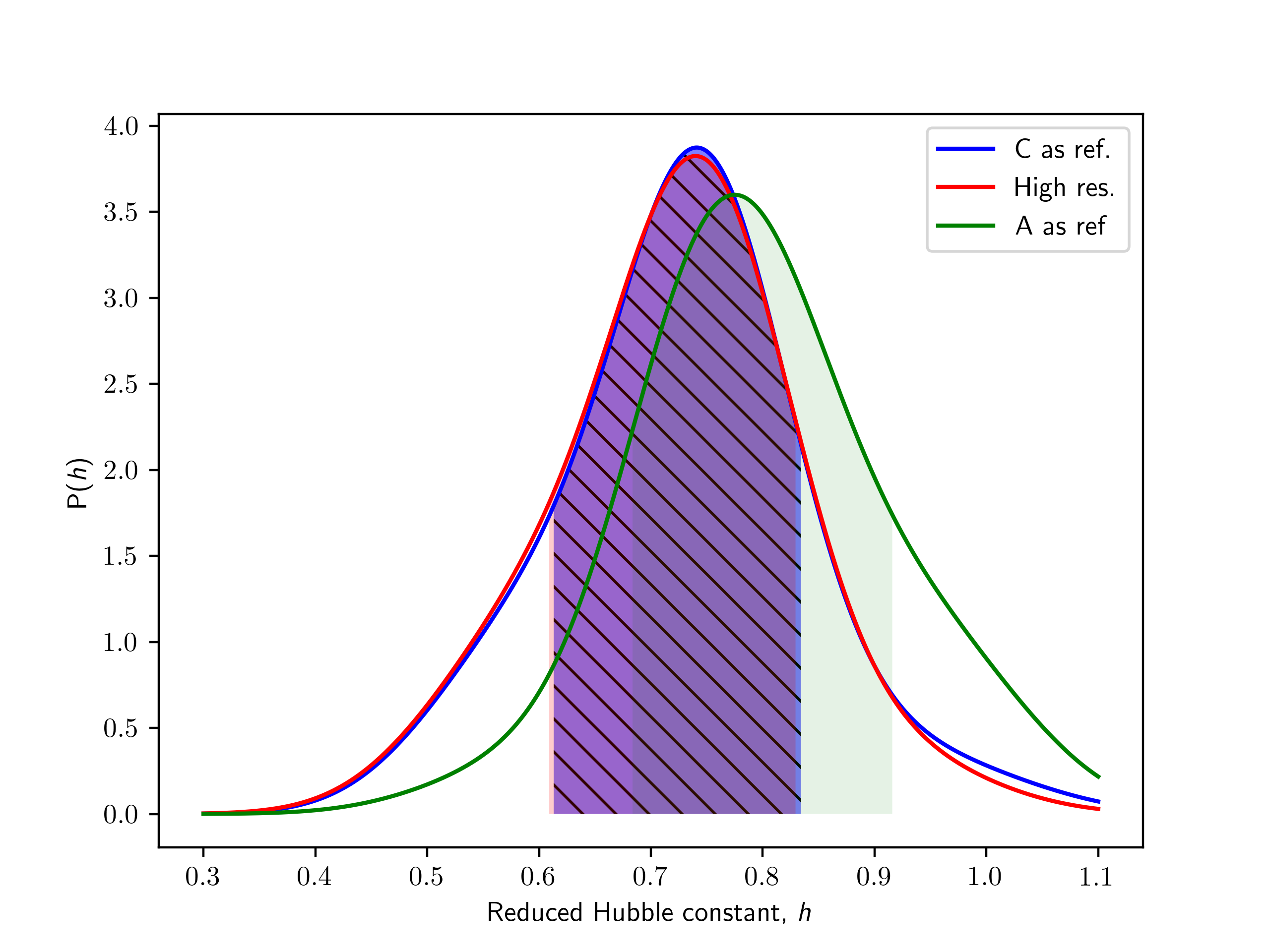}
    \caption{Probability distribution of $h$ obtained from time-delay predicted by the lens models using different analyses.}
    \label{fig:H0Plot}
\end{figure}

Our main results are incorporated in Fig. \ref{fig:H0Plot}. In this figure, we show the probability for the reduced Hubble constant using three different schemes. The standard scheme relies on Eqs. (\ref{Eq_P1}-\ref{eq:weight}) and the 20 lens models, leading to the probability distribution shown in blue. Selecting the 11 lens models with the highest resolutions (248 and 299 grid points), we were also able to obtain the distribution in red. Additionally, we considered the 20 lens models and a modified version of Eq. \eqref{eq:weight}, including flux ratios with respect to A (flux ratios $F_B/F_A$, $F_C/F_A$, and $F_D/F_A$) and the corresponding model magnification ratios, to produce the probability distribution in green. The shaded regions outline the 68\% confidence intervals. Thus, 
\begin{align*}
              H_0 &= 74^{+9}_{-13}\SI{}{\kilo\meter\per\second\per\mega\parsec} \hspace{1cm}\text{(standard scheme)}\\
              H_0 &= 74^{+9}_{-14}\SI{}{\kilo\meter\per\second\per\mega\parsec} \hspace{1cm}\text{(selecting high-resolution} \\
              & \hspace{4.5cm}\text{lens models)}\\
              H_0 &= 77^{+14}_{-11}\SI{}{\kilo\meter\per\second\per\mega\parsec} \hspace{1cm} \text{(including the flux and} \\
              &\hspace{4.5cm} \text{magnifications of the A image)}
\end{align*}

\section{Conclusions}

These results are in agreement with recent estimates reported in \citet{liu2023hubble} and  \citet{napier2023hubble}. The present paper, along with these two works, show that the Hubble constant must lie in the $60-80$ km s$^{-1}$ Mpc$^{-1}$ range. The uncertainty in $H_0$ can be reduced in the future with deeper observations of this cluster, which would reveal more lensed galaxies and their corresponding redshifts. This would take advantage of lensing by galaxy clusters, which typically can have over an order of magnitude more lensing constraints than classic galaxy-QSO lensing. Also, future observations of lensed SNe in galaxy clusters will provide competitive constraints on $H_0$, similar to those obtained with SN Refsdal \citep[][]{Vega2018,Kelly2023}, or SN H0pe \citep{Frye2023}. 

To conclude, we analyze the contribution of this work to the Hubble tension problem. The Hubble tension refers to the 4$\sigma$ to $6\sigma$ discrepancy between estimations of $H_0$ from early universe measurements and late universe results. The most precise predictions of the Hubble constant using early universe results come from the CMB Plank collaboration \citep{collaboration2020planck} that obtained $H_0 = 67.49\pm 0.53$ km s$^{-1}$ Mpc$^{-1}$ with 68\% confidence level (CL). In the late universe, however, the most precise results come from different techniques: using distance ladder Cepheid calibration in the  Large Magellanic, \citet{riess2019large}  estimated $H_0 = 74.03 \pm 1.42$ km s$^{-1}$Mpc$^{-1}$; the SNe Ia distance calibration by the SH0ES collaboration  \citep{riess2022comprehensive} that obtained $H_0 = 73.30 \pm 1.04$ km s$^{-1}$ Mpc$^{-1}$ (68\% CL); and gravitational lensing on quasars by H0LiCOW collaboration \citep{10.1093/mnras/stz3094} that obtained $H_0=73.3^{+1.7}_{-1.8}$ km s$^{-1}$ Mpc$^{-1}$ (68\% CL) and TDCOSMO 
 with $H_0 = 77.1^{+7.3}_{-7.1}$ km s$^{-1}$ Mpc$^{-1}$ \citep{Shajib_2023} in their latest result. There is a chance that the tension might be fictitious (e.g., the resulting effects of the selection of the SNe Ia or use of inadequate lens mass models) or real, in which case a modified cosmological model would be required \citep[e.g.,][]{dainotti2021hubble,Goico, Shalyapin_2023}.  Nonetheless, our results do not show any tension with any of the results in either the early universe or late universe, although this is largely due to the lens model uncertainties. 

\begin{acknowledgements}
We thank Dominique Sluse for his valuable comments and suggestions that allowed us to correct and improve this manuscript. As an answer to one of his suggestions, we incorporated Table \ref{tab:dominique} in the Appendix in the current paper, as these can be useful in microlensing studies.

In addition, Luis J. Goicoechea acknowledges the support from the grant PID2020-118990GB-I00 funded by MCIN/AEI/10.13039/501100011033.
\end{acknowledgements}

\bibliographystyle{aa}
\bibliography{references.bib}

\begin{appendix}
\section{Additional estimated parameters}

Table \ref{tab:dominique} includes the convergence, $\kappa$; shear, $\gamma$; and shear direction, $\phi,$ for the quasar's five images for the 20 lens models used in this work for the estimation of the Hubble constant, together with the mean value for these in the last row of the Table.

\begin{table}[h]
    \centering
\resizebox{\columnwidth}{!}{%
    \begin{tabular}{cc|ccccc}
          &   &QSO.A&QSO.B&QSO.C&  QSO.D &QSO.E\\
          \hline
          \hline
          &   $\kappa$& 0.85&0.71 & 0.70 & 1.09 & 8.19\\
 199/Model1 & $\gamma$& 0.54 & 0.40 & 0.74 & 1.40 & 4.99\\
 & $\phi$& 14.23 & 10.94 & 18.31 & 27.26 & 39.33\\
 \hline
           &   $\kappa$& 0.88 & 0.74 & 0.75 & 1.13 & 8.04\\
 199/Model2 & $\gamma$& 0.56 & 0.42 & 0.75 & 1.40 & 4.89\\
 & $\phi$& 14.55 & 11.28 & 18.41 & 27.21 & 39.22\\
 \hline
           &   $\kappa$& 0.80 & 0.67 & 0.64 & 1.02 & 8.39\\
 199/Model3 & $\gamma$& 0.54 & 0.41 & 0.74 & 1.42 & 5.14\\
 & $\phi$& 14.08 & 11.03 & 18.32 & 27.47 & 39.50\\
 \hline
           &   $\kappa$& 0.86 & 0.73 & 0.73 & 1.11 & 8.09\\
 199/Model4 & $\gamma$& 0.55 & 0.41 & 0.75 & 1.40 & 4.92\\
 & $\phi$& 14.45 & 11.23 & 18.35 & 27.22 & 39.26\\
 \hline
           &   $\kappa$& 0.87 & 0.74 & 0.74 & 1.10 & 7.88\\
 199/Model5 & $\gamma$& 0.56 & 0.42 & 0.75 & 1.39 & 4.81\\
 & $\phi$& 14.64 & 11.48 & 18.43 & 27.15 & 39.13\\
 \hline
           &   $\kappa$& 0.87 & 0.74 & 0.75 & 1.10 & 7.78\\
 199/Model6 & $\gamma$& 0.57 & 0.43 & 0.75 & 1.39 & 4.76\\
 & $\phi$& 14.76 & 11.62 & 18.48 & 27.13 & 39.07\\
 \hline
           &   $\kappa$& 0.87 & 0.75 & 0.75 & 1.10 & 7.71\\
 199/Model7 & $\gamma$& 0.57 & 0.43 & 0.75 & 1.39 & 4.73\\
 & $\phi$& 14.86 & 11.73 & 18.52 & 27.11 & 39.03\\
 \hline
            &   $\kappa$& 0.88 & 0.75 & 0.76 & 1.10 & 7.59\\
 199/Model8 & $\gamma$& 0.58 & 0.44 & 0.76 & 1.39 & 4.67\\
 & $\phi$& 15.00 & 11.88 & 18.56 & 27.09 & 38.95\\
 \hline
           &   $\kappa$& 0.85 & 0.72 & 0.72 & 1.06 & 7.67\\
 199/Model9 & $\gamma$& 0.57 & 0.44 & 0.75 & 1.39 & 4.74\\
 & $\phi$& 14.93 & 11.82 & 18.52 & 27.15 & 39.04\\
 \hline
           &   $\kappa$& 0.86 & 0.74 & 0.72 & 1.03 & 7.01\\
 248/Model1 & $\gamma$& 0.62 & 0.48 & 0.76 & 1.36 & 4.41\\
 & $\phi$& 15.81 & 12.80 & 18.69 & 26.83 & 38.61\\
 \hline
           &   $\kappa$& 0.88 & 0.77 & 0.75 & 1.03 & 6.63\\
 248/Model2 & $\gamma$& 0.63 & 0.49 & 0.76 & 1.33 & 4.19\\
 & $\phi$& 16.14 & 12.97 & 18.58 & 26.49 & 38.28\\
 \hline
           &   $\kappa$& 0.88 & 0.77 & 0.74 & 1.00 & 6.52\\
 248/Model3 & $\gamma$& 0.63 & 0.48 & 0.75 & 1.31 & 4.12\\
 & $\phi$& 16.12 & 12.74 & 18.49 & 26.28 & 38.18\\
 \hline
           &   $\kappa$& 0.85 & 0.73 & 0.72 & 1.01 & 6.99\\
 248/Model4 & $\gamma$& 0.61 & 0.47 & 0.77 & 1.34 & 4.40\\
 & $\phi$& 15.75 & 12.53 & 18.87 & 26.62 & 38.60\\
 \hline
           &   $\kappa$& 0.81 & 0.67 & 0.63 & 0.93 & 7.38\\
 248/Model5 & $\gamma$& 0.64 & 0.48 & 0.75 & 1.30 & 4.07\\
 & $\phi$&15.18 & 12.01 & 18.84 & 26.70 & 38.93\\
 \hline
           &   $\kappa$& 0.88 & 0.76 & 0.72 & 0.99 & 6.39\\
 248/Model6 & $\gamma$& 0.64 & 0.48 & 0.75 & 1.30 & 4.07\\
 & $\phi$&  16.30 & 12.80 & 18.35 & 26.27 & 38.10\\
 \hline
            &   $\kappa$& 0.85 & 0.75 & 0.72 & 1.01 & 7.06\\
 299/Model1 & $\gamma$& 0.61 & 0.47 & 0.75 & 1.33 & 4.41\\
 & $\phi$& 15.63 & 12.53 & 18.44 & 26.58 & 38.62\\
 \hline
           &   $\kappa$& 0.89 & 0.79 & 0.77 & 1.07 & 6.72\\
 299/Model2 & $\gamma$& 0.63 & 0.48 & 0.75 & 1.37 & 4.21\\
 & $\phi$& 16.05 & 12.78 & 18.37 & 26.89 & 38.31\\
 \hline
           &   $\kappa$& 0.80 & 0.67 & 0.63 & 0.92 & 7.36\\
 299/Model3 & $\gamma$& 0.57 & 0.43 & 0.76 & 1.35 & 4.61\\
 & $\phi$& 14.84 & 11.58 & 18.65 & 26.71 & 38.888\\
 \hline
           &   $\kappa$& 0.81 & 0.68 & 0.63 & 0.90 & 7.17\\
 299/Model4 & $\gamma$& 0.58 & 0.43 & 0.76 & 1.33 & 4.50 \\
 & $\phi$& 15.01 & 11.62 & 18.54 & 26.54 & 38.74\\
 \hline
           &   $\kappa$& 0.88 & 0.78 & 0.75 & 1.02 & 6.59\\
 299/Model5 & $\gamma$& 0.63 & 0.48 & 0.74 & 1.32 & 4.17\\
 & $\phi$& 16.07 & 12.88 & 18.31 & 26.47 & 38.25\\
 \hline
            &   $\kappa$& $0.86\pm0.03$ &$0.73\pm0.04$ &$ 0.72\pm0.04 $& $1.04\pm0.06$ & $7.36\pm0.59$\\
 Mean & $\gamma$& $0.59\pm0.03 $ & $0.45\pm0.03$ & $0.75\pm0.01$ & $1.36\pm0.03$ & $4.57\pm0.31$ \\
 & $\phi$ & $15.2\pm0.7$ & $12.0\pm0.7$ & $18.50\pm0.16$ & $26.9\pm0.4$ & $38.8\pm0.4$\\
 
 \end{tabular}
 }
    \caption{Convergence $\kappa$, shear $\gamma$ and shear direction $\phi$ of the multi-resolution lens models for the different QSO images.}
    \label{tab:dominique}
\end{table}
\end{appendix}

\end{document}